# Dipole Response of Spaser on an External Optical Wave

E. S. Andrianov,[1,2] A.A. Pukhov,[1,2] A.V. Dorofeenko,[1,2] A.P. Vinogradov,[1,2] and A.A. Lisyansky[3]

[1]Moscow Institute of Physics and Technology, 9 Institutskiy per., Dolgoprudniy 141700, Moscow Reg., Russia

[2]Institute for Theoretical and Applied Electromagnetics, Izhorskaya 13, Moscow 125412, Russia

[3]Department of Physics, Queens College of the City University of New York, Flushing, NY 11367

Abstract

We find the conditions upon the amplitude and frequency of an external electromagnetic field at which the dipole moment of a Bergman-Stockman spaser oscillates in antiphase with the field. For these values of the amplitude and frequency the losses in metal nanoparticles is exactly compensated by gain. This shows that spasers may be used as inclusions in designing lossless metamaterials.

Interest in the optics of metamaterials – negative index materials – has grown explosively over the last decade (see Refs. [1-3] and references therein). One of the most promising features of metamaterials is the possibility of achieving super-resolution imaging by creating super- and hyper- lenses [4-8] that transform near fields into far fields [4]. Such lenses must be lossless because both loss and gain destroy the image. Though natural metamaterials have not been found, artificial composite materials with desirable properties could be used for this purpose. Metal nanoparticles (NPs) have become the most popular components of such composites. Using surface plasmons (SPs) generated in metal NPs has allowed for the host of innovative applications. The main limiting factor in using metal NPs is their high level of loss [9, 10]. This obstacle can be overcome by introducing a gain medium into the system [11, 12]. The combination of a gain medium and NPs results in the emergence of spasers first suggested by Bergman and Stockman [9]. Exploiting spasers, opens a groundbreaking way to dramatically reduce loss in composites incorporating metal NPs [13, 14].

The spaser cannot be directly used to amplify an external field oscillating with the frequency $\omega_E$ because the spaser is a self-oscillating system and possesses an inherent frequency $\omega_{auto}$, which we refer to below as an autonomous frequency [3, 15]. Nevertheless, as has been shown in Ref. [16], there is a domain (the Arnold tongue) of values of the amplitude $E$ of the external wave and of the frequency detuning $\Delta = \omega_{auto} - \omega_E$ at which the spaser becomes synchronous with the driving wave. Outside of this domain, the spaser exhibits stochastic oscillations. If $\Delta \neq 0$, then there is a threshold value of the external field intensity, $E_{Synch}(\Delta)$, below which the spaser always exhibits stochastic oscillations. The question remains whether within the Arnold tongue the exact cancellation of losses and gain can be achieved.

In this Letter, we study the operation regimes of the spaser driven by an external optical wave. We find a relationship between the amplitude and frequency of the field at which the NP dipole moment oscillates with the phase shift of $\pi$ relatively to the external wave. For the found set of values of $E$ and $\Delta$ the losses at the NP are exactly compensated by gain.

The simplest model of spaser consists of a two-level quantum dot (QD) of the radius $r_{TLS}$, which is positioned at a distance $r$ from a metallic NP of the radius $r_{NP}$. The whole system is immersed into a solid dielectric or semiconductor matrix with dielectric permittivity $\varepsilon_M$ [9]. Below, we discuss the excitation of the main, dipole, mode of the SP at a frequency $\omega_{SP}$.

Following the theory of a usual one-atom single-mode laser [17], the dynamics of the autonomous spaser can be described by the following Hamiltonian:

$$\hat{H}_{auto} = \hbar\omega_{TLS}\hat{\sigma}^\dagger\hat{\sigma} + \hbar\omega_{SP}\hat{a}^\dagger\hat{a} + \hbar\Omega_R(\hat{a}^\dagger\hat{\sigma} + \hat{\sigma}^\dagger\hat{a}),$$

which in the Heisenberg picture leads to the following system of equations (see e.g. [10, 16])

$$\dot{\hat{D}} = 2i\Omega_R(\hat{a}^\dagger\hat{\sigma} - \hat{\sigma}^\dagger\hat{a}) - (\hat{D} - \hat{D}_0)/\tau_D, \qquad (1)$$

$$\dot{\hat{\sigma}} = (i\delta_{TLS} - 1/\tau_\sigma)\hat{\sigma} + i\Omega_R\hat{a}\hat{D}, \qquad (2)$$

$$\dot{\hat{a}} = (i\delta_{SP} - 1/\tau_a)\hat{a} - i\Omega_R\hat{\sigma}, \qquad (3)$$

In Eqs. (1)-(3) we used the rotating wave approximation in which the annihilation operator of the dipole SP and the transition operator $|g\rangle\langle e|$ between ground $|g\rangle$ and excited $|e\rangle$ states of the QD can be represented as $\hat{a}\exp(-i\omega_{auto}t)$ and $\hat{\sigma}\exp(-i\omega_{auto}t)$, respectively. Thus, the operator for the dipole moment of the QD has the form $\hat{\mathbf{d}}_{TLS} = \boldsymbol{\mu}_{TLS}\left(\hat{\sigma}\exp(-i\omega_{auto}t) + \hat{\sigma}^\dagger\exp(i\omega_{auto}t)\right)$, where $\boldsymbol{\mu}_{TLS} = \langle e|e\mathbf{r}|g\rangle$ is the QD dipole moment matrix element; $\Omega_R = \sqrt{3/(r_{NP}\hbar\partial\varepsilon_{NP}/\partial\omega)}(\boldsymbol{\mu}_{TLS}\cdot\mathbf{E}_1)$ is the Rabi frequency with $\mathbf{E}_1 = -\frac{\mathbf{e}}{r^3} + \frac{3(\mathbf{e}\cdot\mathbf{r})\mathbf{r}}{r^5}$ being a field of the unit dipole, $\delta_{TLS} = \omega_{auto} - \omega_{TLS}$ and $\delta_{SP} = \omega_{auto} - \omega_{SP}$. The electric field of the SP is characterized by the NP dipole moment, defined by the operator $\hat{\mathbf{d}}_{NP} = \boldsymbol{\mu}_{NP}\left(\hat{a}\exp(-i\omega_{auto}t) + \hat{a}^\dagger\exp(i\omega_{auto}t)\right)$, where $\boldsymbol{\mu}_{NP} = \sqrt{1.5\hbar r_{NP}^3/(\partial\varepsilon_{NP}/\partial\omega)}\,\mathbf{e}$. The operator $\hat{D} = [\hat{\sigma}^\dagger,\hat{\sigma}] = \hat{n}_e - \hat{n}_g$ describes the population inversion of the ground $\hat{n}_g = |g\rangle\langle g|$ and excited states $\hat{n}_e = |e\rangle\langle e|$ ($\hat{n}_g + \hat{n}_e = \hat{1}$) of the QD, the operator for the population inversion $\hat{D}_0$ describes the pumping [17, 18], the terms with relaxation times $\propto \tau_D^{-1}$, $\tau_\sigma^{-1}$, and $\tau_a^{-1}$ are introduced phenomenologically to account for the relaxation processes of the respective quantities. As has been shown in Refs. [9, 19], above the threshold pump level, $D_{th} = (1 + \delta_{SP}^2\tau_a^2)/(\Omega_R^2\tau_a\tau_\sigma)$, the spaser oscillates with the autonomic frequency $\omega_{auto} = (\omega_{SP}\tau_a + \omega_{TLS}\tau_\sigma)/(\tau_a + \tau_\sigma)$.

Below we neglect quantum fluctuations and correlations and consider $\hat{a}(t)$, $\hat{\sigma}(t)$, and $\hat{D}(t)$ as complex valued quantities (C-numbers), so that we can use complex conjugation instead of the Hermitian conjugation [9, 19-21]. The population inversion $D(t)$ must be a real valued quantity because the respective operator is Hermitian. The quantities $\sigma(t)$ and $a(t)$ are the complex amplitudes of the dipole oscillations of the QD and SP, respectively.

Let us consider the dynamics of the NP and QD forced by the field of the external optical wave $E(t) = E(\exp(i\omega_E t) + \exp(-i\omega_E t))/2$. Assuming that the external field is classical, and taking into account only the dipole interaction, we can write the Hamiltonian in the form

$$\hat{H}_{ef} = \hat{H}_{auto} + \hat{\mathbf{d}}_{NP}\mathbf{E}(t) + \hat{\mathbf{d}}_{TLS}\mathbf{E}(t). \tag{4}$$

As before, we use the Heisenberg equations for the operators $\hat{a}$, $\hat{\sigma}$, and $\hat{D}$ to obtain equations of motion for slowly varying amplitudes:

$$\dot{\hat{D}} = 2i\Omega_R(\hat{a}^\dagger\hat{\sigma} - \hat{\sigma}^\dagger\hat{a}) + 2i\Omega_2(\hat{\sigma} - \hat{\sigma}^\dagger) - (\hat{D} - \hat{D}_0)/\tau_D, \qquad (5)$$

$$\dot{\hat{\sigma}} = (i\delta - 1/\tau_\sigma)\hat{\sigma} + i\Omega_R\hat{a}\hat{D} + i\Omega_2\hat{D}, \qquad (6)$$

$$\dot{\hat{a}} = (i\Delta - 1/\tau_a)\hat{a} - i\Omega_R\hat{\sigma} - i\Omega_1, \qquad (7)$$

where $\Omega_1 = -\mu_{NP}E/\hbar$, $\Omega_2 = -\mu_{TLS}E/\hbar$, and $\delta = \omega_E - \omega_{TLS}$ and $\Delta = \omega_E - \omega_{SP}$ are detuning of frequencies.

It is worth noting that, in the absence of the QD ($\hat{\sigma} = 0$, $\Omega_R = 0$), the stationary solution, $\dot{\hat{a}} = \dot{\hat{\sigma}} = 0$, to Eqs. (5)-(7) gives the following expression for the NP polarization:

$$\alpha/r_{NP}^3 = \frac{3}{2(\partial\varepsilon/\partial\omega)(\Delta + i/\tau_a)}.$$

For small detuning, $\Delta \ll 1$, in the slowly changing amplitudes approximation of, this expression coincides with the classical expression $\alpha^{class} r_{NP}^{-3} = \frac{\varepsilon_{NP}(\omega) - \varepsilon_{ext}}{\varepsilon_{NP}(\omega) + 2\varepsilon_{ext}}$, here $\varepsilon_{ext}$ is the permittivity of the matrix, surrounding the NP.

Without pumping, the frequency dependence of the dipole moment of the NP is similar to that of a classical dipole: its real part may be positive or negative depending on the detuning, $\Delta$, whereas the imaginary part is always positive due to losses. When the pumping is present, the imaginary part of the dipole may change sign and become negative for some values of $\Delta$ (see Fig. 1) signifying the amplification of the incident wave.

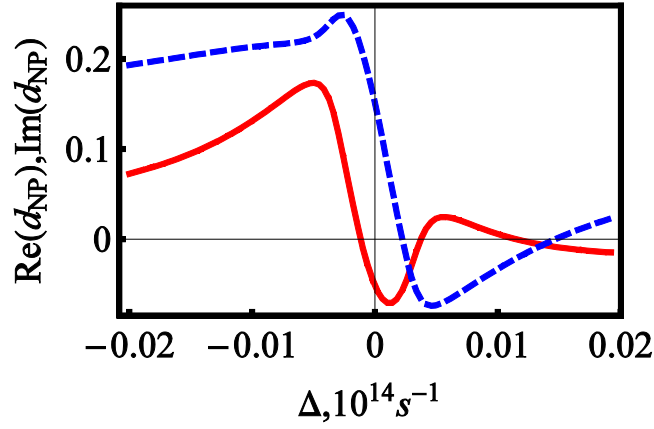

Fig. 1. The dependences of the real (solid line) and imaginary (dashed line) parts of the NP dipole moment on the frequency detuning $\Delta$ for a value of the amplitude of the external field greater than the synchronization threshold, $E > E_{Synch}(\Delta)$.

Figure 1 shows that for fixed $E$ there are two values of $\Delta$ for which an exact compensation of losses can be achieved. These points form a single curve in the $\{E,\Delta\}$ plane (Fig. 2). The equation of this curve $E(\Delta)$ is

$$(\mu_{NP}E/\hbar)^2 = \frac{-(\tau_\sigma^3/\tau_a)\Delta^4 + \left[D_0\Omega\tau_\sigma^3(\mu_{TLS}/\mu_{NP})/\tau_D\right]\Delta^3 - \left[(\tau_\sigma/\tau_D) - \Omega^2 D_0 \tau_a \tau_\sigma^2/\tau_D\right]\Delta^2}{4(\tau_\sigma \Delta \mu_{TLS}/\mu_{NP} + \tau_a \Omega)^2}.$$

As $\Delta \to 0$ this expression transforms into $(\mu_{NP}E/\hbar)^2 = (D_0 - D_{th})\Delta^2 (\tau_\sigma^2/(\tau_D \tau_a))$, where $D_{th} = 1/(\tau_\sigma \tau_a \Omega^2)$ is the threshold pumping. In Fig. 2, where the dipole phase as a function of $E$ and $\Delta$ is shown, the curve $E(\Delta)$ is seen to be a discontinuity line. This line corresponds to a phase difference of $\pi$ between the NP dipole moment and the external field. At this curve, the system has neither loss nor gain, both of which destroy the perfect image of a sub-wavelength lens.

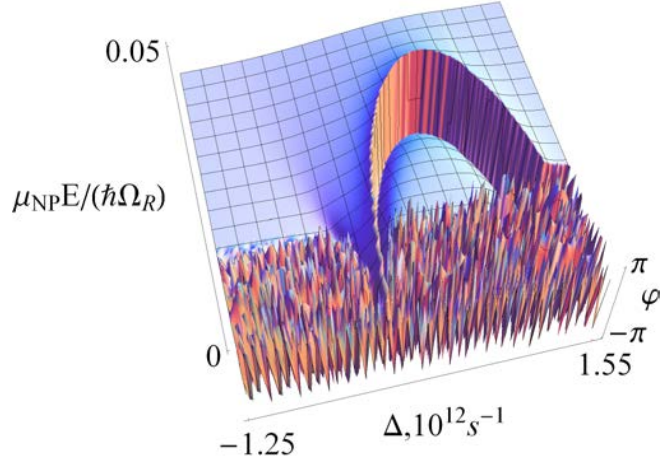

Fig. 2. The dependence of $\phi = \tan^{-1}(\operatorname{Im} d_{NP} / \operatorname{Re} d_{NP})$ on the amplitude of the external field $E$ and the detuning $\Delta$. The smooth part of the surface corresponds to the Arnold tongue where the spaser is synchronized by the external field. At the discontinuity line, on which $\phi = \pi$, the loss is exactly compensated.

The detailed investigation of the dependence of the plasmon dipole moment on the amplitude of the external field for different values of the detuning shows that there are three regimes. For $E < E_{Synch}(\Delta)$, the point $(\Delta, E)$ is outside of the Arnold tongue and the spaser is in the stochastic regime. Inside the Arnold tongue, for weak fields, the amplitude of the dipole moment of the NP depends mainly on pumping and weakly on the amplitude of the external electromagnetic field (Fig. 3). Finally, the linear response regime occurs for very high values of the electric field, $E > 0.5 \times 10^6$ V/m, when the coupling of the NP with external field becomes comparable to the coupling between the NP and the QD.

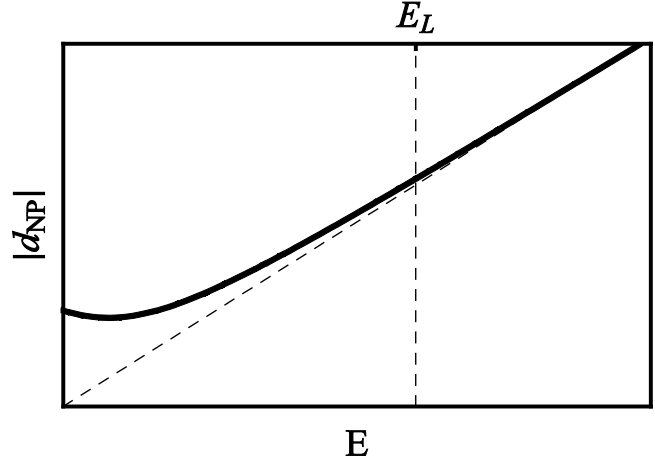

Fig. 3. The dependence of the plasmon dipole moment on the amplitude of the external field for zero frequency detuning ($\Delta = 0$).

While the composite, represented by a matrix filled with spasers, may amplify or weaken the external field in the general case, there is a combination of the amplitude and frequency of the external field that provides exact cancellation of losses and gain and results in antiphase oscillations of the spaser dipole moment. Such a composite would be a novel nonlinear medium, which may follow the frequency of the controlling signal. The properties of the medium may differ from the properties of a single spaser due to possible mutual synchronization. We plan to study this problem in the near future.

The authors are grateful to J. Pendry for critical discussion. This study was partly supported by RBRF Grants No 10-02-92115, 10-02-91750 and PSC-CUNY grant

[1] W. Cai and V. Shalaev, *Optical Metamaterials* (Springer, 2010).
[2] R. Marques, F. Martin, and M. Sorolla, *Metamaterials with negative parameters: theory, design and microwave applications* (Wiley, 2008).
[3] M. Premaratne and G. P. Agrawal, *Light propagation in gain medium* (Cambridge University Press, 2011).
[4] J. B. Pendry, Phys. Rev. Lett. **85**, 3966 (2000).
[5] P. A. Belov and Y. Hao, Phys. Rev. B **73**, 113110 (2006).
[6] P. A. Belov, C. R. Simovski, and P. Ikonen, Phys. Rev. B **71**, 193105 (2005).
[7] P. A. Belov, Y. Zhao, S. Tse, P. Ikonen, M. G. Silveirinha, C. R. Simovski, S. Tretyakov, Y. Hao, and C. Parini, Phys. Rev. B **77**, 193108 (2008).
[8] Z. Liu, H. Lee, Y. Xiong, C. Sun, and X. Zhang, Science **315**, 1686 (2007).

[9] D. J. Bergman and M. I. Stockman, Phys. Rev. Lett. **90**, 027402 (2003).
[10] I. E. Protsenko, A. V. Uskov, K. E. Krotova, and E. P. O'Reilly, J. Phys.: Conf. Ser. **107**, 012010 (2008).
[11] U. Leonhardt, IEEE J. Sel. Top. Quantum Electron. **9**, 102 (2003).
[12] V. G. Veselago, Sov. Phys. Usp. **10**, 509 (1968).
[13] K. Li, X. Li, M. I. Stockman, and D.J. Bergman, Phys. Rev. B **71**, 115409 (2005).
[14] M. A. Noginov, G. Zhu, A. M. Belgrave, R. Bakker, V. M. Shalaev, E. E. Narimanov, S. Stout, E. Herz, T. Suteewong, and U. Wiesner, Nature **460**, 1110 (2009).
[15] M. I. Stockman, J. Opt. **12**, 024004 (2010).
[16] E. S. Andrianov, A. A. Pukhov, A.V. Dorofeenko, A. P. Vinogradov, and A. A. Lisyansky, arXiv:1105.5656.
[17] M. O. Scully and M. S. Zubairy, *Quantum optics* (Cambridge University Press, 1997).
[18] R. H. Pantell and H. E. Puthoff, *Fundamentals of quantum electronics* (Wile, 1969).
[19] I. E. Protsenko, Phys. Rev. A **71**, 063812 (2005).
[20] A. N. Lagarkov, A.K. Sarychev, V.N. Kissel, and G. Tartakovsky, Phys. Usp. **52**, 959 (2009).
[21] A. S. Rosenthal and T. Ghanan, Phys. Rev. A **79**, 043824 (2009).